\documentclass[aps,pra,twocolumn,superscriptaddress,amsmath,amssymb,floatfix,showpacs]{revtex4}

\usepackage[dvips]{graphicx}

\usepackage{amsfonts}
\usepackage{amssymb}
\usepackage{amsmath}

\begin{document}

\title{Dynamic and Energetic Stabilization of
Persistent Currents in Bose-Einstein Condensates}
\author{K.J.H.\ Law}
\affiliation{Computer, Electrical and Mathematical Sciences \& Engineering,
King Abdullah University of Science and Technology, Thuwal 23955-6900, KSA}
\author{T.W.\ Neely}
\altaffiliation[Current address: ]{School of Mathematics and Physics, University of Queensland, Qld 4072, Australia}
\affiliation{College of Optical Sciences, University of Arizona,
  Tucson, AZ, 85721, USA}
\author{P.G.\ Kevrekidis}
\affiliation{Department of Mathematics and Statistics,
University of Massachusetts,
Amherst MA 01003-4515, USA}
\author{B.P.\ Anderson}
\affiliation{College of Optical Sciences and Department of Physics,
University of Arizona, Tucson, AZ, 85721, USA.}
\author{A.S.\ Bradley}
\affiliation{Jack Dodd Centre for Quantum Technology, Department of
Physics, University of Otago, P. O. Box 56, Dunedin, New Zealand}
\author{R.\ Carretero-Gonz{\'a}lez}
\affiliation{%
Nonlinear Dynamical Systems Group,\footnote{{\tt URL:} http://nlds.sdsu.edu/}
Department of Mathematics and Statistics, 
and Computational Science Research Center, 
San Diego State University, San Diego, CA, 92182-7720, USA. }


\begin{abstract}

We study conditions under which vortices in a highly oblate harmonically 
trapped Bose-Einstein condensate (BEC) can be stabilized due to 
pinning by a blue-detuned Gaussian laser beam, with particular emphasis on the potentially destabilizing effects of laser beam positioning within the BEC.  Our approach involves theoretical and numerical exploration of dynamically and energetically stable pinning of vortices with winding number up to $S=6$, in correspondence with experimental observations.  
Stable pinning is quantified theoretically via Bogoliubov-de Gennes
excitation spectrum computations and confirmed via direct numerical simulations for a range of conditions similar to those of experimental observations.   The theoretical and numerical results indicate that the pinned winding number, or equivalently the winding number of the superfluid current about the laser beam, decays as a laser beam of fixed intensity moves away from the BEC center.  Our theoretical analysis helps explain previous experimental observations, and helps define limits of stable vortex pinning for future experiments involving vortex manipulation by laser beams.

\end{abstract}

\maketitle

\section{Introduction}

The persistence of superfluid flow and superconducting currents about barriers, and the related topic of the pinning of quantized vortices and magnetic flux, appear as signature phenomena of superfluidity and superconductivity~\cite{tilley}.  While there have been numerous experimental investigations on quantized vortices in atomic Bose-Einstein condensates (BECs)~\cite{fetter2009,Anderson2010}, relatively few experiments have explored parameters for which one or more barriers within a BEC can localize and inhibit the motion of singly or multiply quantized vortices.  Evidence for such vortex pinning by laser beams was demonstrated in studies of the interactions of vortex lattices in rotating BECs and co-rotating optical lattices~\cite{Sch2004.PRL93.210403}, as well as with single laser beams piercing the BEC~\cite{Weiler2008,Samson2012,vortex_manipulation}.  Experiments involving the persistence of superfluid flow about a single laser barrier centered within a BEC extend the concept of vortex pinning and determine timescales over which superfluid flow can be maintained in annular traps.  In these experiments, macroscopic superfluid flow in annular traps has been obtained from internal atomic state manipulation~\cite{ryu2007a,Ram2011.PRL106.130401,Beattie2013}, weak-link rotation~\cite{Wright2013}, laser path engineering~\cite{Samson2012}, and the decay of two-dimensional quantum turbulence~\cite{Neely2013}.  Theoretically and numerically, stable vortex pinning about a central potential barrier has been examined \cite{Kuo2010.PRA81.033627,Kuo2010.JLTP161.561} for various heights and widths of the potential, and for BECs with few atoms or weak interatomic interactions.  However, for larger or more strongly interacting BECs, and especially regarding the influence of other parameters such as beam position on the stability of superfluid flow and vortex pinning~\cite{Samson2012,vortex_manipulation}, the stability problem has not been fully explored.   Furthermore, as the field continues to evolve, vortex state engineering methods utilizing vortex pinning and manipulation are becoming more feasible~\cite{Samson2012}, and there is an increasing need to better understand conditions for which vortices can be stably pinned and manipulated within a BEC.

In this article, motivated by recent experimental observations \cite{ryu2007a,Neely2013} that suggest persistent current lifetimes in BECs may be limited by the position of a pinning laser beam, and by new methods for generating and manipulating vortices in BECs~\cite{Samson2012}, we theoretically and numerically explore the dynamical and energetic stability of vortex pinning about a laser barrier.  In our approach, we consider the laser intensity, width, and position within a two-dimensional (2D) harmonically trapped BEC.  Our physical scenario corresponds to the parameters of
Ref.~\cite{Neely2013} in which superfluid flow in a highly oblate BEC was established about a laser beam through the decay of 2D quantum turbulence. We also present new experimental observations suggesting that the decay of net superfluid flow may be in part due to laser beam position drift, similar to the conclusions of Ref.~\cite{ryu2007a}. Our main result is that for the trap and BEC parameters associated with the experimental observations, the number of vortices that can be pinned by the beam drops as the beam intensity decreases or as the pinning potential moves away from the center of the BEC, consistent with the experimental observations. Our results additionally suggest that in currently developing methods involving vortex manipulation in BECs, regimes of stable pinning must be considered for the engineering of persistent currents and complex vortex or pinning site distributions.

Our discussion is structured as follows. To set up and motivate the theoretical and numerical problem, we first discuss general concepts and experimental observations of the decay of superfluid flow in an annular, highly oblate BEC, and include experimental evidence for the drift in the position of the beam relative to the trap center as the BEC is held in the trap.   Following this, we describe the model setup and theoretical background that supports our analysis, and present our main computational results. Finally, we summarize our findings and discuss directions for future study. A brief appendix provides details regarding our numerical methods.


\section{General Concepts and Experimental Motivation}

Vortices of topological charge $S=1$ are dynamically stable in BECs.  They are routinely observed in experiments and their dynamics can be accordingly followed~\cite{Nee2010.PRL104.160401,dshall}.  
However, such single quantized
vortices represent excited states of the system, a feature
that especially
at non-zero temperatures in a stationary harmonic trap,
has significant dynamic implications.
In a highly oblate BEC, dissipation due to interaction of the vortices with thermal excitations will cause the vortices to spiral out of the trap, or vortices of opposite circulation to
annihilate one another and convert their energy into acoustic energy within the BEC.  Vortices are thus not inherently energetically stable.

In the case of higher charge ($S>1$) vortices, energetic instability is accompanied by an intrinsic dynamical instability that was originally evaluated in Ref.~\cite{pu}; see also Refs.~\cite{Kuo2010.PRA81.033627,Kuo2010.PRA81.023603,kollar} for recent detailed mathematical analyses of this instability that favors the splitting of multi-charge vortices into single-charge vortices with the same total winding number. This, in turn, renders preferable the loose clustering of multiple vortices rather than their perfect co-location.  With the addition of a blue-detuned laser beam that pierces a BEC, theoretical analysis has shown that singly and multiply quantized vortices can be stably pinned by the laser
beam~\cite{Kuo2010.PRA81.033627,Kuo2010.JLTP161.561,vortex_manipulation}.  As a general concept for axially symmetric systems, as long as the vortex remains pinned to the beam, and no other vortices are introduced into or leave the system, metastable superfluid flow about the barrier will persist.

The first experimental study of BEC persistent currents involved the creation of superfluid flow about a blue-detuned Gaussian laser beam that acted as a vortex pinning potential within the BEC~\cite{ryu2007a}.  In this study, the authors noted that the lifetime of superflow about the central barrier was limited by drift in the relative positions of the center of the harmonic potential and the initially co-located laser beam. This is an indication of the need to better understand and control the parameters involved in the stability of vortex pinning and superflow persistence.  In the experiment of Ref.~\cite{Neely2013} on two-dimensional quantum turbulence, similar effects were observed; although not discussed in Ref.~\cite{Neely2013}, the results of these observations are presented below, and serve as a primary motivation for our theoretical investigation.   Recent experimental methods have now minimized or eliminated such relative drift and enabled superflow to persist for up to 2 minutes ~\cite{Beattie2013}.

We base our study on the experiment of Ref.~\cite{Neely2013}.  In this experiment, a blue-detuned laser was used to stir vortices into highly oblate BECs of $\sim 2\times 10^6$ $^{87}$Rb atoms held in a trap with radial ($r$) and axial ($z$) trapping frequencies of $(\omega_r/2\pi,\,\omega_z/2\pi) = (8,\,90)$ Hz.  After the 0.33~s stir, and an equilibration period lasting 1.66~s, the BEC is held in the annular trap for a variable hold time $t_h$; in the following discussion, $t_h=0$ corresponds to the end of the equilibration period.  At the beginning of the hold period, the system is at a temperature $T\sim47$~nK, and the BEC critical temperature at this point is $\sim 82$~nK.  Once the hold period ends, the central barrier is ramped off over 0.25~s, the trapping potential is removed, and the BEC ballistically expands for $\sim$50~ms and is then observed using standard absorption imaging techniques.   For the cases in which a persistent current about the central barrier exists prior to barrier ramp-down and expansion, a hole was observed in the expanded density distribution; the area of the core can be used to determine the winding number of superfluid flow around the central barrier~\cite{Murray2013}.    Alternatively, the BEC may be held for an additional 3~s in the trap after the barrier ramp off but before the expansion imaging procedure, allowing the vortices to separate and become experimentally distinguishable.  In this case, the number of vortices observed corresponds to the winding number of the current that was pinned prior to barrier ramp-down.  Both methods were used to determine the winding number of the persistent current as a function of $t_h$.  Similar experimental techniques for observing vortex pinning, superfluid currents, and winding number have been used in numerous experiments; see for example Refs.~\cite{ryu2007a,Weiler2008,Mou2013.PRA86.013629,Murray2013}.

\begin{figure}[htb]
\includegraphics[width=1\columnwidth]{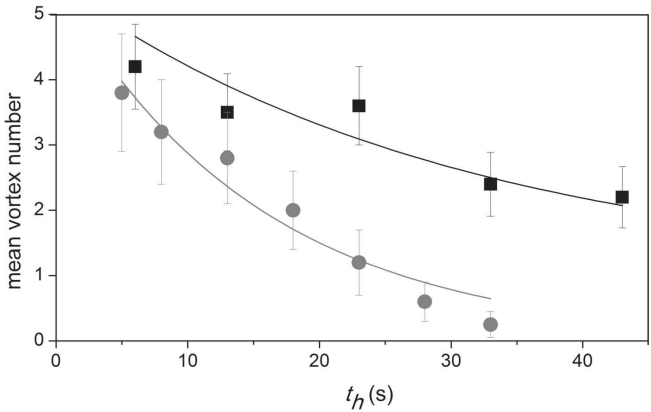}
\caption{\label{figS2}
Evidence of superfluid current persistence, and its decay.  Black squares show the mean number of vortices formed and held in the presence of the laser barrier, observed in 10 runs for each of the hold times $t_h$ shown, with 3~s of hold time used to separate the vortices for determining the vortex number.
Error bars represent statistical uncertainty.  An exponential fit to the
data (black curve) gives a $1/e$ lifetime of 31(4)~s.  Gray circles show
the lifetime of vortices in the system under the same conditions, but with
the pinning beam ramped down at the beginning of the hold time
rather than at the end.  An exponential fit (gray curve) gives the lifetime
of free vortices to be 15(1)~s, an indication that vortex and current lifetime decreases without the central barrier present.  The BEC lifetime decreases with a $1/e$ lifetime of 24(3)~s (not shown).}
\end{figure}

\begin{figure}[htb]
\includegraphics[width=.85\columnwidth]{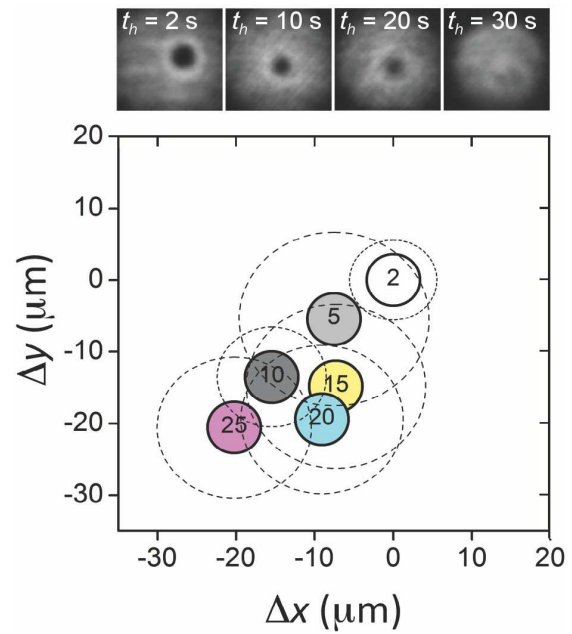}
\caption{\label{drift_figure}
Experimental indication of relative drift between the harmonic trap center and pinning laser beam. (top row)  Each image of expanded BECs, acquired with a procedure described in the text, is an average of 5 images taken under identical conditions and for the hold times indicated.  For $t_h=2$ s, a large fluid-free core indicating pinned vorticity is observed on the right side of the BEC center; this core is presumed to originate at the repeatable initial position of the pinning beam, prior to any drift.  The darkness of this core in the BEC after image averaging indicates that this position is consistent from shot to shot.  For later hold times, a consistent drift in the core position relative to its initial position is observed.  For $t_h=30$ s, the core is much less visible, and the relative position of the pinning beam and the harmonic trap is therefore much less repeatable from shot to shot.  (bottom panel) Colored circles indicate the changes $\Delta x$ and $\Delta y$ of the position of the core in expanded images, relative to the mean core position at $t_h=2$ s, for the hold times shown in the centers of each circle (times given in seconds).  The magnitude of statistical uncertainty is indicated with larger dashed circles about the center of each mean position.  Position uncertainties for $t_h=30$ s and later times are unusually large compared with earlier times, and are thus not shown.  As in the top row of images, data are obtained from averages over 5 experimental runs at each hold time given.  }
\end{figure}

As described and shown in Refs.~\cite{Neely2013,Rooney2013}, annular superflow corresponding to a winding number up to $S\sim 5$ was created in this stirring and equilibration procedure, and was observed to persist for $1/e$ times on the order of 30 s, as indicated in Fig.~\ref{figS2}.  For this plot, the mean number of vortices was counted after decay of the persistent current into individual vortices in the harmonic trap after barrier ramp-down, as described above.  Each black square represents the mean vortex number 
observed in 10 runs.  The time-dependent drop in vortex number can be caused by any mechanism that destabilizes a vortex pinned to the central barrier, allowing it to spiral out of the BEC, although such vortex dynamics were not directly confirmed in the experiment.  If the central barrier was instead removed at the {\em beginning} of the hold period, immediately after equilibration, and the system was left to evolve in a purely harmonic trap, the mean number of vortices observed dropped at a faster rate.  These data are indicated in Fig.~\ref{figS2} by gray circles.  The higher number of vortices observed in the cases where the central barrier was present is an indication of superflow being maintained by the Gaussian barrier beam.  

One likely mechanism for the observed decay of the supercurrent is relative drift of the pinning beam with respect to the harmonic trap center, as was earlier speculated in Ref.~\cite{ryu2007a}.  In order to investigate any such drift in the experiment of Ref.~\cite{Neely2013}, the BEC was expanded immediately after barrier ramp-down, and the density hole due to the pinned vorticity was thus visible.  By direct averaging of 5 images taken under identical conditions, for various hold times, the relative position of the dark density hole with respect to the center of the fitted Thomas-Fermi BEC profile was studied. The results are summarized in Fig.~\ref{drift_figure}.  As observed in the figure, for the first 25 s of hold, a relative drift of the pinning beam can be observed in the data, and serves as a possible mechanism for the decay of superfluid current indicated in Fig.~\ref{figS2}.  This relative drift could result from beam drift due to movement by the laser mirrors, or due to changes in the magnetic trap position due to a changing temperature of the magnetic field coil.

These experimental observations serve as the primary motivating factor for the theoretical study presented in the remainder of this paper.  A full set of numerical simulations of vortex dynamics under a wide range of possible experimental systematic errors is beyond the scope of this study.  Rather, our study is designed to address the question of energetic and dynamical stability of vortex pinning from a theoretical standpoint, and including the potentially destabilizing effects of an off-center Gaussian pinning potential.


\section{Theoretical Setup}

In order to explore the above observations from a theoretical perspective, we start our considerations from a non-dimensional grand-canonical energy functional for a BEC in the mean-field approximation~\cite{pethickpit} of the form:
\begin{equation}
H = \int d{\bf r'}\,[\, |\nabla \Psi)|^{2}
+ V({\mathbf r'}) |\Psi|^2 +
\frac{1}{2} |\Psi|^4 - \mu |\Psi|^2 \,]
\label{hamiltonian}
\end{equation}
\noindent where $d\mathbf{r'}$ is a volume element, $\Psi({\mathbf r}') \in \mathbb{C}$ is the BEC order parameter at 3D position $\mathbf{r'}=(x,y,z)$, and $\mu$ is the chemical potential associated with the conservation of the number of atoms $N = \int d{\bf r'} |\Psi|^2$.
Defining $\mathbf{r}=(x,y)$, $V({\mathbf r'}) \in \mathbb{R}$ is the external confining potential, of 
the form:
\begin{equation}
V({\mathbf r}, z) = \underbrace{
\left(\frac{1}{2}|{\mathbf r}|\right)^2 +
\left(\frac{\omega_z}{2 \omega_r}      z      \right)^2
}_{V_{{\rm T}}}+
\underbrace{V_0 \,
e^{\frac{-2 |{\bf r}-{\bf r}_0|^2}{w^2}}
}_{V_{{\rm B}}},
\label{potential}
\end{equation}
\noindent where $V_{{\rm T}}$ is the trap's parabolic confining potential with
fixed angular frequencies matching those described in the previous section.
creating a highly oblate potential amenable to a two-dimensional (2D) reduction used in the remainder of this paper.  
$V_{{\rm B}}$ is the repulsive potential due to a blue-detuned laser-beam potential of peak barrier energy $V_0$ and Gaussian radius $w$ centered
at 2D position ${\bf r}_0$ relative to the center of the harmonic trap $V_T$, enabling stability analysis for both on-center and off-center beams.
The time, energy, and length scales in Eqs.~(\ref{hamiltonian}) and (\ref{potential}) are, respectively,
$1/\omega_r$, $\hbar \omega_r$, and $\sqrt{\hbar/2 m\omega_r}$,
where $m$ is the atomic mass.
For $^{87}$Rb with an $s$-wave scattering
length of $a_s=5.5$~nm, this amounts to the dimensionless $N$
being connected to the experimentally measured atom number through
a multiplicative factor of $N_{{\rm fac},2D}\simeq 20$. From here on, when we
refer to $N$, this multiplicative conversion will be implied and
the symbol will stand for the experimentally relevant atom number.

The resulting equation of motion for this
Hamiltonian system
$(\dot\Psi,c.c.)^T=J \sigma (\delta
H/\delta\Psi,c.c.)^T=J \sigma D H$,
is the Gross-Pitaevskii equation~\cite{pethickpit},
where $D$ is the functional gradient of $H(\Psi,\Psi^*)$,
$J={\rm diag}(-iI,iI)$ with $I$ the identity operator
and $\sigma$ interchanges row $2$ with
$1$.
The GPE can be 
written as:
\begin{equation}
i \dot{\Psi} = - \nabla^2 \Psi + V({\mathbf r}) \Psi + |\Psi|^2 \Psi
- \mu \Psi.
\label{gpe}
\end{equation}
Of particular importance for our considerations in what follows will be
the stability of stationary solutions $\Psi$ to
Eq.~(\ref{gpe}). This is determined by the
eigenvalues $\{\epsilon\}$ of the Hessian of the Hamiltonian,
$ {\bf H} \equiv \sigma D^2 H(\Psi) $,
and excitation spectrum $\{\lambda\}$
of the resulting linearization operator $J {\bf H}$ which
corresponds to Bogoliubov-de Gennes
analysis~\cite{pethickpit}.
When considering the excitation spectrum
of single or multi-vortex states, we find in it the existence of
negative energy modes~\cite{skryabin}.
Negative energy eigenvalues (also referred to as anomalous
modes) of
${\bf H}$ indicate {\it energetic instability}, since ``dissipative''
perturbations (e.g.,~from exchanges of atoms with the thermal cloud if the
temperature deviates from zero) in the system can render them {\it dynamically
  unstable}, as can collisions with other eigendirections having
positive energy even in the purely
Hamiltonian (zero-temperature) system.
Nevertheless, in the latter case, energetic instability of an excited
state such as a dark soliton or a vortex~\cite{pethickpit} does
not necessarily lead to dynamical instability.
Thus these modes reveal the
potential of such an excited state towards genuine dynamical instability
which arises in both of the above mentioned scenarios.
The linear stability/excitation spectrum of the
system is monitored through the eigenvalues $\lambda=\lambda_r + i \lambda_i$
of $J \sigma D^2 H$; dynamical instability arises
when $\lambda_r\neq 0$ since, due to
the Hamiltonian structure, the eigenvalues feature a four-fold
symmetry over the real
and imaginary axes.  From prior experience (see, e.g., Ref.~\cite{yankody}),
which is confirmed
again here, linear stability generically
indicates evolutionary {\it nonlinear (orbital) stability}
in the mean-field model, at least for time scales monitored
of the order of tens of
seconds, i.e., comparable to the lifetime of the BEC.

An additional key observation is that
the number of negative energy modes
depends on the topological charge, or winding number $S$ of the
multi-vortex configuration considered~\cite{kollar}.
In particular, $S$ is mathematically defined as
\begin{eqnarray*}
S=\frac{1}{2\pi}\int_{\partial\Omega} v d {\bf r},~~~~~
v = \nabla\phi,
\end{eqnarray*}
\noindent where $\partial\Omega$ is the boundary of a region
containing the vortices, $v$ 
is the superfluid velocity of the condensate and $\phi$ is the phase
associated with the complex valued wavefunction $\Psi=\sqrt{\rho} e^{i \phi}$.
To examine the role of small amplitude excitations to a stationary
vorticity bearing solution $\Psi=\Psi(r) e^{i S \theta}$
\footnote{For simplicity this is written for a vortex centered
at the origin, although it can be straightforwardly generalized
for a vortex centered at $(r_0,\theta_0)$}, where $\theta$ is the polar angle,
perturbations of the
form $\psi = a({\mathbf r}) e^{\lambda t} + b({\mathbf r})
e^{\lambda^* t}$ are introduced. Given the cylindrical symmetry of the problem,
these can be selected to have a decomposition in polar coordinates given by
$a(r,\theta) = \tilde{\alpha}(r) e^{i \kappa_{a}\theta}$ and
$b(r,\theta) = \tilde{\beta}(r) e^{i \kappa_{b}\theta}$.
In particular, if we set $\kappa_{a}=\kappa$ then $\kappa_{b}=\kappa-2S$,
so a single index $\kappa$ will dictate the angular dependence
of the excitation with given eigenvalue $\lambda$.
Hence, the spectrum of eigenvalues $\{\lambda\}$ can be
decomposed as the union of the spectra $\{\lambda_{\kappa}\}$ pertaining
to perturbations of index $\kappa$.

It has been shown numerically \cite{pu} and
analytically \cite{kollar} that instability windows arise  in
a solution with topological charge $S$ only for indices
%
$|\kappa| < S$.
The null eigenvalues corresponding to the phase or gauge [U(1)]
invariance of the GPE model appear in the spectrum
of $\kappa=S$.  For a vortex in a harmonic trap, there is an
anomalous mode for $\kappa=S-1$ that is typically not resonant with
any modes of positive energy, accounting for the dynamical stability
of the $S=1$ vortex discussed earlier. We also note that
this mode converges to zero as $\omega_r \rightarrow 0$,
restoring translational invariance in the limit
and leading to the energetic stability of the $S=1$
vortex without external potential~\cite{middelkamp}.  For each $0 \leq \kappa < S-1$ in the case of
$S>1$, an anomalous mode can lead to windows of instability, as has
been shown, e.g., in Refs.~\cite{pu,kollar}.  One of the principal
points of the present work is to illustrate that for experimental parameters in use in current experiments, such
anomalous modes can be completely
{\it suppressed} for a strong enough Gaussian pinning potential $V_{\rm B}$, even if the pinning potential is not centered at the center of the harmonic trap.
Indeed, the modes not only cease to resonate with positive modes and
no longer lead to dynamical instability windows, but for even
higher laser powers (and hence larger $V_0$), they undergo a
transition from negative to positive energy.  This precludes the existence
of any such resonances and lends energetic stability to the
corresponding multi-charge vortex state.

An important comment is that
such energetic stability  enhances the observability of
these higher charge configurations.
This is contrary to the experimental difficulties in observing
such higher charge vortices in the absence of the barrier
considered herein, as reported earlier in Ref.~\cite{S2Ket}.
The theoretical explanation of such enhanced observability has
to do with the fact that, as mathematically proved in Ref.~\cite{sandst}
and numerically illustrated, e.g., for the case of single charge
vortices in Ref.~\cite{middelkamp2}, the presence of arbitrarily small
dissipative perturbations immediately destabilizes these anomalous
modes. The role of dissipative perturbations in the case of BECs
is played by the lossy coupling to thermal atoms (in any realistic
finite temperature setting). As a result, vortices develop complex
excitation frequencies in their spectra~\cite{middelkamp2} that lead to their
well-known spiraling out of the condensate; see, e.g., Ref.~\cite{proukakis}.
This effect is
stronger for higher charge vortices, as the latter bear more
anomalous modes as indicated above. Hence, the energetic
stabilization of these anomalous modes (in the absence of dissipation)
``shields'' the multi-charge state from such a detrimental effect.

\section{Computational Results}

\begin{figure}[tbh]
\begin{center}
\includegraphics[width=7.5cm]{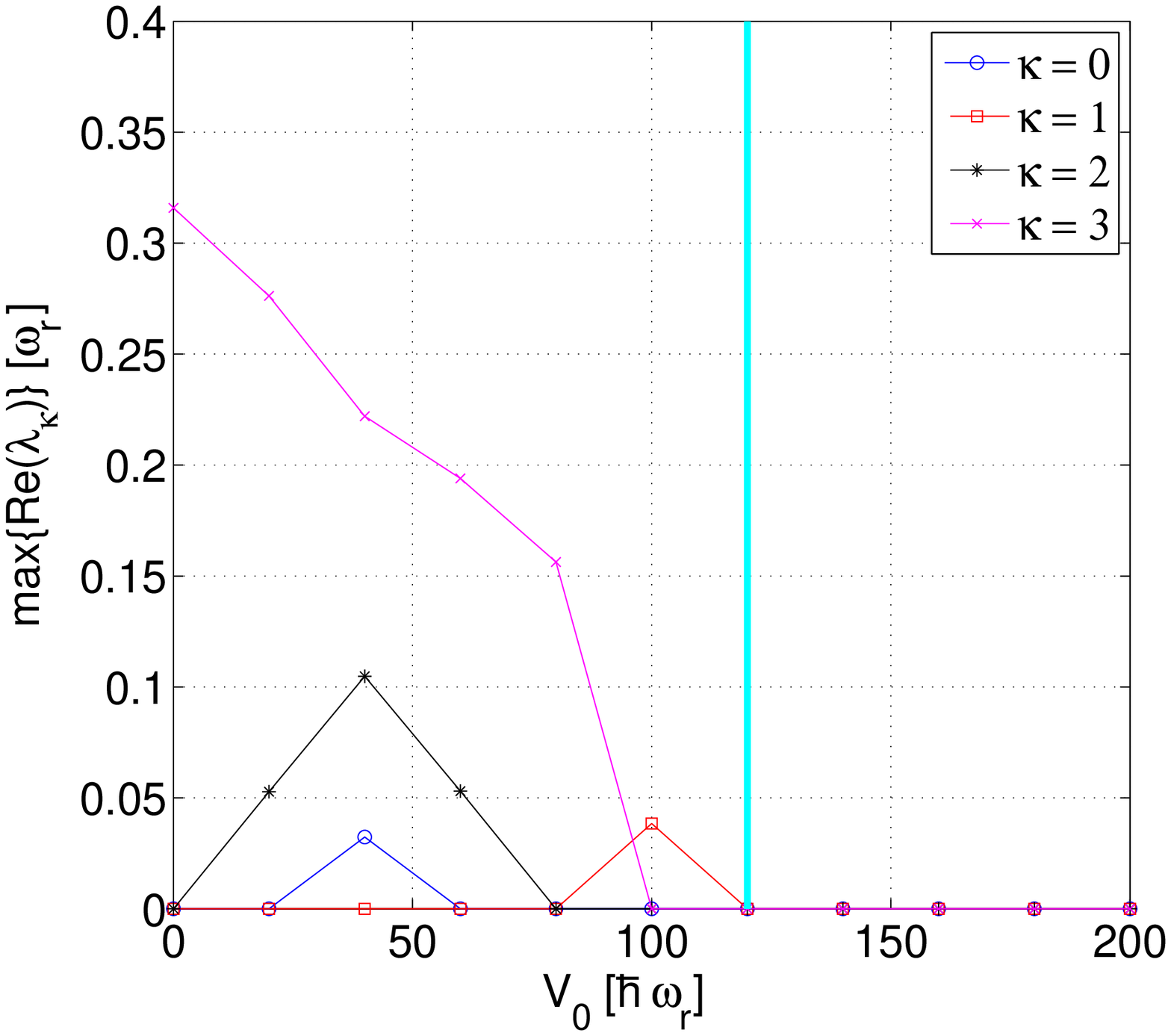}
\includegraphics[width=7.5cm]{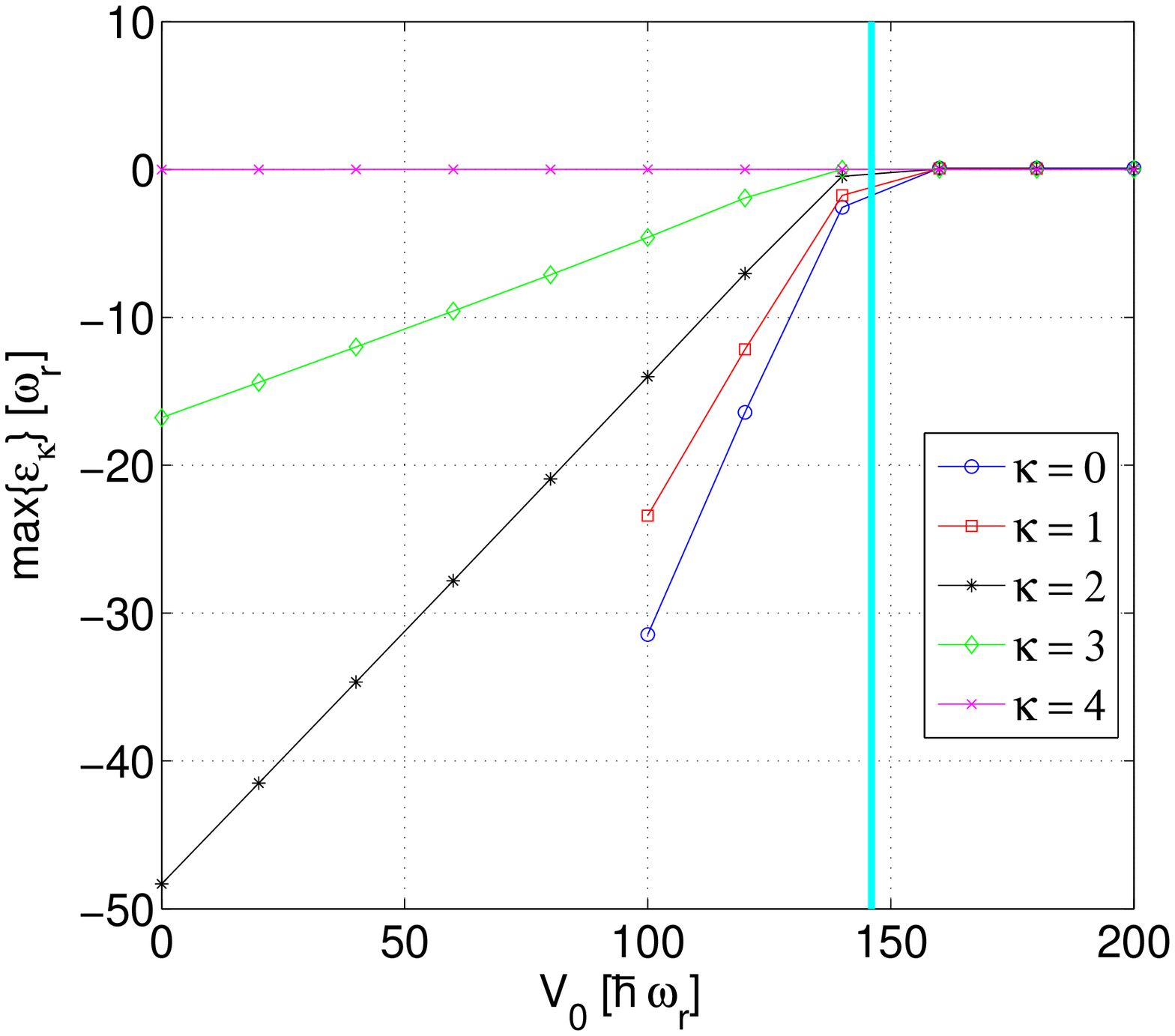}
\end{center}
\caption{%
Maximal growth rate associated to energetically unstable 
index $\kappa$
for a vortex of charge $S=5$, with $w=10.5$ and $N=2\times 10^6$.
Notice the threshold of
dynamical stability associated with the vertical line
in the top panel and the energetic
stability indicated by the vertical line in the bottom panel
(the energy spectra was computed on a finer discretization of $V_0$
in order to more precisely determine
the threshold between the plotted data points).}
\label{fig1}
\end{figure}

\begin{figure}[htb]
\begin{center}
\includegraphics[height=6cm]{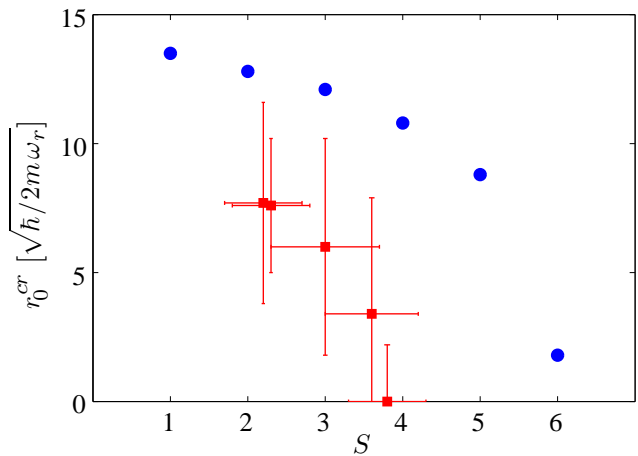}
\end{center}
\caption{Energetic stability threshold radii ($r_0^{cr}$) 
for various charge-$S$ currents pinned by the external laser.
The (blue) circles depict the energetic stability 
obtained from the Bogoliubov-de Gennes analysis
(the resolution in our numerics for $r_0^{cr}$ is $\pm 0.1$).  
The (red) squares depict experimental distances at which
the indicated persistent current charges were observed.  Error bars represent statistical uncertainty of the measurement results.  
All simulations are performed with
($N$,$V_0$, $w$, $\omega_r$, $\omega_z$, $a$, $m$)
corresponding to dimensional parameters ($2\times 10^6$, $146 \hbar \omega_r$, $20 \mu {\rm m}$,
$2 \pi \times 8 {\rm Hz}$,
$2 \pi \times 90 {\rm Hz}$, $5.5 \mu m$, $87 {\rm amu})$.
}
\label{offcenter}
\end{figure}

Figure~\ref{fig1} is a principal example showing,
for experimentally accessible parameters,
the suppression of the instability for a vortex of charge $S=5$
as the pinning laser power ($V_0$) increases.
Since the only potentially unstable modes satisfy $\kappa<S-1=4$,
only indices up to $\kappa=3$ are considered in the top panel
of the figure.  
It can be seen that for
high enough values of $V_0$ (see Fig.~\ref{fig1} for
details), dynamical instability by any of the potentially unstable
perturbation indices $\kappa$
is completely suppressed. Perhaps even more importantly, as illustrated
by the bottom panel, for $V_0 > 146$ all negative energy modes have been
converted to positive energy ones, thus converting the configuration
into an energetically stable one and a likely candidate for
experimental observation.

\begin{table}
\begin{center}
\begin{tabular}{|c|c||c|c||c|c| }
    \hline
  \multicolumn{2}{|c||} {$N{=}2{\times}10^6\!,\,w{=}10.5$} &
  \multicolumn{2}{|c||} {$N{=}2{\times}10^6\!,\,S{=}5$} &
  \multicolumn{2}{|c|} {$w{=}10.5, S{=}5$} \\
  \hline
\hline			
  $S$ & $V_0 [\hbar \omega_r]$ & $w [\sqrt{\hbar/2m\omega_r}]$ &
  $V_0 [\hbar \omega_r]$ & $N$ & $V_0 [\hbar \omega_r]$\\
  \hline \hline
  3 & 142 & 1 &240 &$1.5 {\times}10^6$ & 130\\ \hline
  4 & 144 & 5 &140 & ${\bf 2 {\times}10^6}$ &{\bf 146} \\ \hline
  {\bf 5} & {\bf 146} & {\bf 10.5}& {\bf 146}& $2.5 {\times}10^6$ &162\\ \hline
  6 & 150 & 15 & 162& & \\  \hline
  7 & 154 & & & & \\ \hline
  \end{tabular}
\end{center}
\caption{The value of the threshold $V_0$ for energetic stability
indicated by the vertical in the bottom panel of Fig.~\ref{fig1},
for various charges, $S$ (left), beam waists $w$ (center), and number of atoms $N$ (right).
Bold values correspond to the vertical line in the bottom panel of Fig.~\ref{fig1}, for comparison.
The chemical potential of the BEC with $2\times 10^6$ atoms, $w=10.5$ and $V_0=146$ is
$\mu \approx 90$ in our dimensionless units.}
\label{higher}
\end{table}

A more detailed evaluation of the critical $V_0$ beyond which the
vortices of different topological charge become energetically
stable is given in Table I. The left column shows the increasing
trend of $V_0$ (i.e., higher laser intensities are needed to stabilize higher charge vortices) over $S$, for fixed
width $w$ of the barrier and number of atoms $N$.
The next two columns focus on the specific
case of $S=5$ and again fixed $N$
and monitor the {\it non-monotonic} dependence of
the critical $V_0$, as the width $w$ of the barrier is increasing.
Finally, for fixed charge $S$ and width $w$, once again it is intuitively
expected (and shown in the rightmost two columns of the table) that for stabilization,
the maximum energy of the pinning potential needs to be higher for larger numbers
of atoms in the BEC.
These diagnostics yield a sense of the dependencies
of the vortex stabilization barrier parameters, as the beam width, charge,
and atom number in the BEC are varied.

We now turn to the case where the barrier is not centered
at ${\bf r}_0=0$, but rather at a finite distance $r_0 > 0$
from the center, where we let ${\bf r}_0=(r_0,0)$.
For such an off-center pinning beam, the results
are modified as follows. As illustrated in Fig.~\ref{offcenter},
a stable vortex for a centered beam
becomes unstable at some critical radius $r_0^{cr}$
as the beam is moved outward. This radius, depends on
the intensity of the beam and is found to be larger for fixed
intensity as the charge $S$ decreases.  Hence, for example, an energetically stable
$S=6$ vortex ($V_0=146 \hbar \omega_r$, $w=20 \mu m$, $N=2 \times 10^6$)
will become energetically unstable outside radius $r_0^{cr}=1.8$ while the $S=5$ vortex
for the same parameters will remain energetically stable until a radius
$r_0^{cr}=8.8$.
Finally, in the case of
$S=1$, we find that $r_0^{cr}=13.5$, which essentially implies that this
 state is always stable as it is transferred off center by the barrier,
all the way to the boundary of the BEC.

Our theoretical analysis is also consistent with experimental observations.  The red squares shown in Fig.~\ref{offcenter} show the mean position of the observed persistent current locus (obtained without extra hold time after barrier ramp-down, as in Fig.~\ref{drift_figure}) plotted against mean observed winding number.   The comparison of experimental data and analytically obtained stability limits does not directly show that beam drift is responsible for the decay of the current.  However, the experimental data are contained within (and show
a similar trend as) the theoretically determined stability limits, lending support to the predictions of the theoretical results.  If the beam drift is primarily responsible for the decay of the current, the comparison between 
theoretical and experimental data suggests 
that it is possible that other factors such as thermal fluctuations may 
need to be considered for accurate prediction of persistent current stability 
conditions.

\begin{figure}[t]
\includegraphics[width=8.75cm]{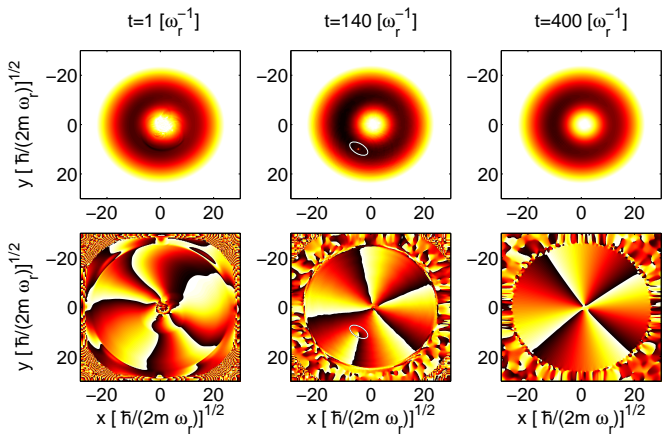}
\caption{Snapshots of the density (top row) and phase (bottom row)
of the evolution of a perturbed off-center $S=5$ state in the regime of
energetic stability for the $S=4$ state
($w=10.5$, $N=2\times 10^6$, and $V_0=144$).
The dissipatively perturbed GPE followed
here dynamically destabilizes the $S=5$ state by ejecting
one vortex (see vortex inside the ellipse in the middle
panels) and
subsequently (and for the duration of the dynamical evolution)
locks into the energetically stable $S=4$ state.
}
\label{dynamics5}
\end{figure}

Consider a centered beam in between the regimes of energetic stability for
successive charge numbers, when charge $S$ is energetically stable
but charge $S+1$ is energetically unstable.  In this parametric
regime, the $S+1$ solution will typically
only be dynamically stable.
When such an on-center solution is shifted
just slightly off-center (i.e., perturbed), the energetic instability manifests
immediately, and it becomes dynamically unstable if the dynamics
followed is dissipative (due to the instability of the remaining
anomalous mode as explained above).  Such a setting
will still be in the energetic stability regime for the
charge $S$ solution, since
energetic stability is robust to such small shifts of symmetry, as
displayed in Fig.~\ref{offcenter}.
To highlight this behavior we introduce phenomenological
dissipation by using the {\it dissipatively perturbed} GPE
\begin{equation}
(i+\gamma) \dot{\Psi} = - \nabla^2 \Psi + V({\mathbf r}) \Psi + |\Psi|^2 \Psi
- \mu \Psi,
\label{dgpe}
\end{equation}
where the coefficient $\gamma$ accounts for the dissipation rate
due to the coupling between the BEC cloud and the non-condensed
(thermal) cloud.
The dissipative GPE has been shown to reliably exhibit the dissipative
effects of coupling with the thermal atoms \cite{dgpe1,dgpe2,dgpe3,dgpe4}.
In our present study, the precise value of the dissipation
$\gamma$, and its relation with the temperature of the
BEC, is not relevant since {\em any} amount of dissipation
will destabilize a dynamically stable but energetically
{\em unstable} solution.
Therefore, we take a nominal value of $\gamma=0.001$ and monitor
the evolution of a dynamically stable but energetically unstable $S+1$
under the dissipative GPE.
One such example is depicted in Fig.~\ref{dynamics5} where an initial stationary
off-centered $S=5$ state, which is originally dynamically stable under the
Hamiltonian (non-dissipative) GPE, is rendered dynamically
unstable by adding dissipation ($\gamma>0$).
The instability is manifested by the ejection of one of the vortices
(see vortex inside the ellipse in the middle panels) that
was originally trapped by the laser. After this vortex is ejected
and absorbed at the edge of the BEC cloud, the remaining $S=4$
state, which is energetically stable for these parameter values,
persists for as long as the dynamical evolution is followed.
This dynamical example corroborates our existence and excitation spectrum
analysis given above and illustrates the significance of the energetic
stability criteria given for the experimental observability of the
different multi-charge configurations.

\section{Conclusions and Future Challenges}

In the present work, we described the results of experiments in which
vorticity with winding number up to $S=5$ can be pinned to a laser beam
in a BEC.  As observed, this topological charge decreases with time,
presumably due to unobserved depinning of a vortex followed by migration
out of the trapping beam in possible combination with vortex-antivortex
annihilation.  Our central question has been to theoretically study the
stability of vortex pinning under our experimental conditions, including
an examination of pinning stability due to off-center beams.
As the beam strength increases, our numerical simulations
illustrate that multi-charge states become progressively
dynamically stabilized (i.e., no complex-valued eigenfrequencies)
and subsequently energetically stabilized (i.e., no anomalous modes).
Hence, there is no potential for instability even
in the presence of dissipative dynamics due to the coupling with
the thermal cloud. The dependence of the relevant anomalous modes
was highlighted not only
as a function of the laser intensity, but also of its width,
the atom number in the condensate and the vortex topological
charge. This was done both for the case where the beam was
centered at the center of the parabolic confinement, as well as for
the case where it was off-center.

There are numerous directions that are opening up for future extensions
of the present considerations. Perhaps the most notable one
is to extend the relevant considerations computationally in
three-dimensional settings  and use them as a way of obtaining
persistent currents associated with vortex lines, or perhaps with more
complex three-dimensional
configurations carrying vorticity. An additional subject
that the present work facilitates which is of intense recent interest
is that of the study of few vortex cluster configurations and their
interactions; see, e.g., Ref.~\cite{dshall2}. The potential ramp down of
the Gaussian beam may provide ideal conditions for the examination
of the multi-charge dynamics and interactions. 

{\it Acknowledgments}. PGK, KJHL and RCG gratefully acknowledge the
support of NSF-DMS-0806762. 
PGK and RCG also acknowledge support from NSF-DMS-1312856.
PGK also acknowledges support from
the Alexander von Humboldt Foundation, the Binational Science 
Foundation under grant 2010239, NSF-CMMI-1000337, 
FP7, Marie Curie Actions, People, International Research
Staff Exchange Scheme (IRSES-606096)
and from  the US-AFOSR under grant FA9550-12-10332.  
TWN and BPA acknowledge the support of NSF through grants PHY-0855467 and PHY-1205713.  
KJHL acknowledges the support of EPSRC, ERC, ONR, ESA, and SRI.  
ASB acknowledges the support of a Rutherford Discovery Fellowship administered by the Royal Society of New Zealand.

\appendix
\section{Numerical Methods}

Our methods extend those in, e.g.,
Refs.~\cite{huepe,todd,law}.
The spatial discretization in $(r,\theta)$ employs Chebyshev polynomials to
represent $r$ dependence~\cite{trefethen}.  The Fourier modes used to represent the
$\theta$ dependence make the Laplacian operator diagonal in this direction.
To identify stationary states of Eq.~(\ref{gpe}), we first obtain an
initial estimate via imaginary-time (i.e., replacing $t \rightarrow it$)
integration using a first-order implicit/explicit Euler scheme with $\Delta t
= 10^{-2}$.  We refine the solution obtained from relaxation
using Newton's method.  The linear system arising at each Newton
step is solved using the matrix-free IDR(s) algorithm
\cite{sonneveld,gijzen},
which requires only the action of the Hessian. To accelerate
inversion, we precondition the system with the inverse Laplacian,
making use of its block diagonal structure.  Hence, we solve the system
$\nabla^{-2} D^2 H(\Psi_n) \Delta_n = \nabla^{-2} D H(\Psi_n)$ and update
$\Psi_{n+1}=\Psi_n-\Delta_n$ for $n=0,1,\ldots$.
Fewer than 5 Newton iterations usually achieve an accuracy of
$||\nabla^{-2}D H(\Psi)||_{l^2}/||\Psi||_{l^2}<10^{-12}$.

For each stationary solution $\Psi$, we use the Implicitly Restarted Arnoldi
algorithm to iteratively compute the eigenpairs of the linearization $J \sigma
D^2 H(\Psi)$ to a specified tolerance \cite{sorenson}.  
In order
to find the desired eigenvalues we use inverse iteration, with
the IDR(s) method and
inverse Laplacian preconditioning to solve the linear systems, as above.
Here, the preconditioner is taken to be $[J \sigma (\nabla^2)]^{-1}$, so that
each iteration solves $\nabla^{-2}D^2 H(\Psi) v_{n+1} = -\nabla^{-2} \sigma J
v_n$.

Finally, the dynamical evolution is carried out
using a standard fourth order Runge-Kutta method
in time and finite differences in space.


\begin{thebibliography}{999}

\bibitem{tilley} D.R. Tilley and J. Tilley,
{\it Superfluidity and Superconductivity}, IOP Publishing (Philadelphia,
1990).

\bibitem{fetter2009} A.L. Fetter,
Rev. Mod. Phys. {\bf 81}, 647 (2009).

\bibitem{Anderson2010}
B.P. Anderson, J. Low Temp. Phys. {\bf 161}, 574 (2010).

\bibitem{Sch2004.PRL93.210403}
V. Schweikhard, I. Coddington, P. Engels, S. Tung,  and E.A. Cornell,
Phys. Rev. Lett. {\bf 93}, 210403 (2004).

\bibitem{Weiler2008}
C.N. Weiler, T.W. Neely, D.R. Scherer, A.S. Bradley, M.J. Davis, and B.P. Anderson, Nature {\bf 455}, 948 (2008).

\bibitem{Samson2012}
E.C. Samson, {\it Generating and Manipulating Quantized Vortices in Highly Oblate Bose-Einstein Condensates},
Ph.D. thesis, University of Arizona (2012).

\bibitem{vortex_manipulation}
M.C. Davis, R. Carretero-Gonz{\'a}lez,
Z. Shi, K.J.H. Law, P.G. Kevrekidis, B.P. Anderson,
Phys. Rev. A {\bf 80}, 023604 (2009).




\bibitem{ryu2007a}
 C. Ryu, M.F. Andersen, P. Clade, V. Natarajan, K. Helmerson, and W.D. Phillips, Phys. Rev. Lett. {\bf 99},
260401 (2007).

\bibitem{Ram2011.PRL106.130401}
A. Ramanathan, K.C. Wright, S.R. Muniz, M. Zelan, W.T. Hill, C.J. Lobb, K. Helmerson, W.D. Phillips, and G.K. Campbell, Phys. Rev. Lett. {\bf 106}, 130401 (2011).

\bibitem{Beattie2013}
S. Beattie, S. Moulder, R.J. Fletcher, and Z. Hadzibabic, Phys. Rev. Lett. {\bf 110}, 025301 (2013).

\bibitem{Wright2013}
K.C. Wright, R.B. Blakestad, C.J. Lobb, W.D. Phillips, and G.K. Campbell, Phys. Rev. Lett. {\bf 110}, 025302 (2013).

\bibitem{Neely2013}
T.W. Neely, A.S. Bradley, E.C. Samson, S.J. Rooney, E.M. Wright, K.J.H. Law, R. Carretero-Gonz{\'a}lez, P.G. Kevrekidis, M.J. Davis,  and B.P. Anderson, Phys. Rev. Lett. {\bf 111}, 235301 (2013).


\bibitem{Kuo2010.PRA81.033627}
P. Kuopanportti and M. M\"{o}tt\"{o}nen, Phys. Rev. A {\bf 81}, 033627 (2010).

\bibitem{Kuo2010.JLTP161.561}
P. Kuopanportti and M. M\"{o}tt\"{o}nen, J. Low Temp. Phys. {\bf 161}, 561 (2010).

\bibitem{Nee2010.PRL104.160401}
T.W. Neely, E.C. Samson, A.S. Bradley, M.J. Davis, and B.P. Anderson, Phys. Rev. Lett. {\bf 104}, 160401 (2010).

\bibitem{dshall} D.V. Freilich, D.M. Bianchi, A.M. Kaufman, T.K. Langin, and D.S. Hall, Science {\bf 329}, 1182 (2010).


\bibitem{pu} H. Pu, C.K. Law, J.H. Eberly and N.P. Bigelow,
Phys. Rev. A {\bf 59}, 1533 (1999).


\bibitem{Kuo2010.PRA81.023603}
P. Kuopanportti, E. Lundh, J.A.M. Huhtam\"{a}ki, V. Pietil\"{a}, and M. M\"{o}tt\"{o}nen, Phys. Rev. A {\bf 81}, 023603 (2010).

\bibitem{kollar} R. Koll{\'a}r and R.L. Pego,
App. Math. Res. eXpress {\bf 2012}, 1 (2012).


\bibitem{Murray2013}
N. Murray, M. Krygier, M. Edwards, K.C. Wright, G.K. Campbell, and C.W. Clark, Phys. Rev. A {\bf 88}, 053615 (2013).

\bibitem{Rooney2013}
S.J. Rooney, T.W. Neely, B.P. Anderson, and A.S. Bradley, Phys. Rev. A {\bf 88}, 063620 (2013).



\bibitem{Mou2013.PRA86.013629}
S. Moulder, S. Beattie, R.P. Smith, N. Tammuz, and Z. Hadzibabic, Phys. Rev. A {\bf 86},  013629 (2012).


\bibitem{pethickpit} C.J. Pethick and H. Smith,
{\it Bose-Einstein condensation in dilute gases}, Cambridge University
Press (Cambridge, 2002).
L.P. Pitaevskii and S. Stringari,
{\it Bose-Einstein Condensation}, Oxford University Press (Oxford, 2003).


\bibitem{skryabin} 
see for a detailed discussion of such modes: 
D.V. Skryabin, Phys. Rev. A {\bf 63}, 013602 (2000).

\bibitem{yankody} K.J.H. Law, L. Qiao, P.G. Kevrekidis, and I.G. Kevrekidis
Phys. Rev. A {\bf 77}, 053612 (2008).


\bibitem{middelkamp} S. Middelkamp, P.G. Kevrekidis, D.J. Frantzeskakis,
R. Carretero-Gonz{\'a}lez and P. Schmelcher, Phys. Rev. A {\bf 82}, 013646
(2010).

\bibitem{S2Ket}
A.E. Leanhardt, A. G\"orlitz, A.P. Chikkatur, D. Kielpinski, 
Y. Shin, D.E. Pritchard, and W. Ketterle,
Phys. Rev. Lett. {\bf 89}, 190403 (2002);
%
Y.~Shin, M.~Saba, M.~Vengalattore, T.A.~Pasquini, C.~Sanner,
A.E.~Leanhardt, M.~Prentiss, D.E.~Pritchard, and W.~Ketterle,
Phys.\ Rev.\ Lett.\ {\bf 93}, 160406 (2004).

\bibitem{sandst} T. Kapitula, P.G. Kevrekidis and B. Sandstede,
Physica D {\bf 195}, 263 (2004).

\bibitem{middelkamp2} S. Middelkamp, P.G. Kevrekidis, D.J. Frantzeskakis,
R. Carretero-Gonz{\'a}lez and P. Schmelcher, J. Phys. B {\bf 43}, 155303
(2010).

\bibitem{proukakis} A.J. Allen, E. Zaremba, C.F. Barenghi, N.P. Proukakis,
Phys. Rev. A {\bf 87}, 013630 (2013).

\bibitem{dgpe1}
M. Tsubota, K. Kasamatsu, and M. Ueda,
Phys. Rev. A {\bf 65} 023603 (2002).

\bibitem{dgpe2}
K. Kasamatsu, M. Tsubota, and M. Ueda,
Phys. Rev. A {\bf 67}, 033610 (2003).

\bibitem{dgpe3}
S. Choi, S.A. Morgan, and K. Burnett,
Phys. Rev. A {\bf 57}, 4057 (1998).

\bibitem{dgpe4}
S.J. Rooney, T.W. Neely, B.P. Anderson, and A.S. Bradley,
Phys. Rev. A {\bf 88}, 063620 (2013).


\bibitem{dshall2} R. Navarro, R. Carretero-Gonz{\'a}lez, P.J. Torres, 
P.G. Kevrekidis, D.J. Frantzeskakis, M.W. Ray, E. Altunta\c{s}, D.S. Hall,
Phys. Rev. Lett. {\bf 110}, 225301 (2013).

\bibitem{huepe} C. Huepe, L.S. Tuckerman, S. Metens, M.E. Brachet,
Phys. Rev. A \textbf{68}, 023609 (2003).

\bibitem{todd} T. Kapitula, K.J.H. Law, P.G. Kevrekidis,
SIAM J. Appl. Dyn. Syst. {\bf 9}, 34 (2010).

\bibitem{law}
K.J.H. Law, P.G. Kevrekidis, and L.S. Tuckerman,
Phy Rev Lett. {\bf 105}, 160405 (2010).


\bibitem{trefethen} L.N. Trefethen, {\it Spectral Methods in MATLAB},
  SIAM (Philadelphia, 2000).

\bibitem{sonneveld} Peter Sonneveld and Martin B. van Gijzen,
SIAM J. Sci. Comput. Vol. 31, No. 2, pp. 1035-1062 (2008).


\bibitem{gijzen} http://ta.twi.tudelft.nl/NW/users/gijzen/IDR.html

\bibitem{sorenson} http://www.caam.rice.edu/software/ARPACK/


\end{thebibliography}
\end{document}